\documentclass[10pt,conference]{IEEEtran}
\usepackage{amsmath,amssymb,amsfonts}

\usepackage{hyperref}
\usepackage{cleveref}
\usepackage{ifluatex}
\usepackage[normalem]{ulem}
\usepackage{cite}
\usepackage[xindy,style=index,numberedsection=false]{glossaries}
\usepackage{tikz}
\usepackage{ifthen}
\usepackage{circledsteps}
\usepackage[]{algorithm,algpseudocode}
\usepackage{graphicx}
\usepackage{textcomp}
\usepackage{subfigure}
\usepackage{xcolor}
\usepackage{xurl}
\usepackage{booktabs}
\usepackage{makecell}
\usepackage{caption}
\usepackage[binary-units=true]{siunitx} 
\usepackage[bb=dsserif]{mathalpha}
\def\BibTeX{{\rm B\kern-.05em{\sc i\kern-.025em b}\kern-.08em
    T\kern-.1667em\lower.7ex\hbox{E}\kern-.125emX}}

\usepackage{setspace}
\let\Algorithm\algorithm
\renewcommand\algorithm[1][]{\Algorithm[#1]\setstretch{1.1}}
\algnewcommand{\LineComment}[1]{\State \(\triangleright\) #1}
\algtext*{EndWhile} 
\algtext*{EndIf} 
\algtext*{EndFor} 
\algtext*{EndFunction} 

\newacronym{sae}{SAE}{Sparse Approximate Eigenproblem}
\newacronym{evd}{EVD}{Eigenvalue Decomposition}
\newacronym{ff}{FF}{Flip Flop}
\newacronym{lut}{LUT}{Look Up Tables}
\newacronym{sa}{SA}{Systolic Array}
\newacronym{nlp}{NLP}{Natural Language Processing}
\newacronym{ppr}{PPR}{Personalized PageRank}
\newacronym{spmv}{SpMV}{Sparse Matrix-Vector Multiplication}
\newacronym{topkspmv}{Top-K SpMV}{Top-K Sparse matrix-vector multiplication}
\newacronym{coo}{COO}{Coordinate}
\newacronym{dsl}{DSL}{Domain-Specific Language}
\newacronym{slr}{SLR}{Super Logic Region}
\newacronym{csc}{CSC}{Compressed Sparse Column}
\newacronym{csr}{CSR}{Compressed Sparse Row}
\newacronym{raw}{RAW}{Read-After-Write}
\newacronym{vram}{VRAM}{Video RAM}
\newacronym{ndcg}{NDCG}{Normalized Discounted Cumulative Gain}
\newacronym{ir}{IR}{Information Retrieval}
\newacronym{dcg}{DCG}{Discounted Cumulative Gain}
\newacronym{er}{ER}{Entity Resolution}
\newacronym{ii}{II}{Initiation Interval}
\newacronym{gpu}{GPU}{Graphics Processing Unit}
\newacronym{fpga}{FPGA}{Field Programmable Gate Array}
\newacronym{colamd}{COLAMD}{ColumnApproximate Minimum Degree algorithm}
\newacronym{cuda}{CUDA}{Compute Unified Device Architecture}
\newacronym{cpu}{CPU}{Central Processing Unit}
\newacronym{hbm}{HBM}{High Bandwidth Memory}
\newacronym{ddr}{DDR}{Double Data Rate}
\newacronym{bscsr}{BS-CSR}{Block-Streaming CSR}
\newacronym{glove}{GloVe}{Global Vectors for Word Representation}
\newacronym{cu}{CU}{Compute Unit}
\newacronym{pe}{PE}{Processing Element}
\newacronym{sp}{SP}{Stream Processor}
\newacronym{uram}{URAM}{UltraRAM}
\newacronym{cordic}{CORDIC}{Coordinate Rotation Digital Computer}
\newacronym{iram}{IRAM}{Implicitly Restarted Arnoldi Method}
\newacronym{dsa}{DSA}{Domain Specific Architecture}

\newcommand{\GB}[1]{\SI{#1}{\giga\byte}}

\newcommand{\lanczos}{Lanczos}

\ifluatex
    
\else
    
\fi

\begin{document}
\title{A Mixed Precision, Multi-GPU Design for Large-scale Top-K Sparse Eigenproblems}

\author{
\IEEEauthorblockN{
Francesco Sgherzi\IEEEauthorrefmark{1},
Alberto Parravicini\IEEEauthorrefmark{2},
Marco D. Santambrogio\IEEEauthorrefmark{2}}
\IEEEauthorblockA{Politecnico di Milano, DEIB, Milan, Italy}
\IEEEauthorrefmark{1}{francesco1.sgherzi}@mail.polimi.it
\IEEEauthorrefmark{2}{\{alberto.parravicini, marco.santambrogio\}}@polimi.it,
\vspace{20pt}
}
\maketitle

\begin{abstract}
Graph analytics techniques based on spectral methods process extremely large sparse matrices with millions or even billions of non-zero values. 
Behind these algorithms lies the Top-K sparse eigenproblem, the computation of the largest eigenvalues and their associated eigenvectors.
In this work, we leverage GPUs to scale the Top-K sparse eigenproblem to bigger matrices than previously achieved while also providing state-of-the-art execution times.
We can transparently partition the computation across multiple GPUs, process out-of-core matrices, and tune precision and execution time using mixed-precision floating-point arithmetic. 
Overall, we are 67$\times$ faster than the highly optimized ARPACK library running on a 104-thread CPU and 1.9$\times$ than a recent FPGA hardware design.
We also determine how mixed-precision floating-point arithmetic improves execution time by 50\,\% over double-precision, and is 12$\times$ more accurate than single-precision floating-point arithmetic.
\end{abstract}


\section{Introduction}\label{sec:intro}
\glsunset{cuda}
Modern data science and numerical mathematics applications operate on larger and larger data, often with strict requirements of minimizing execution time and power consumption.
For many of these applications, hardware accelerators such as \glspl{gpu} and \glspl{fpga} are a highly-effective solution, especially when mixed-precision and reduced-precision arithmetic come into play~\cite{clark2010solving,gupta2015deep,clark2018pushing,sze2020efficient,parravicini2020reduced,sgherzi2021solving}. 
As spectral methods become ubiquitous in the large scale graph pipelines of Spectral Clustering \cite{ng2002spectral}, \gls{ir} \cite{langville2005survey} and ranking \cite{dawson2021opinionrank}, such techniques require algorithms that can compute only a subset of the most relevant eigenvalues (i.e. the largest in modulo) and their associated eigenvectors while taking advantage of the sparsity of real-world graphs.
Graph analytics pipelines usually operate on graphs with millions or even billions of edges, rendering traditional methods computing all eigenvalues impractical, as their space and time cost scales at least quadratically with the number of vertices.
As the size of real-world graphs exceeds the device memory size of modern \glspl{gpu}, an eigensolver must be able to handle out-of-core matrices as well as be capable of distributing the computation to multiple \glspl{gpu}. 
Moreover, they need to support different numerical data types for storage and computation to meet storage and accuracy requirements.

In this work, we introduce a novel Top-K \gls{gpu} eigensolver for real-valued, sparse matrices capable of handling matrices with billion of non-zero entries, of partitioning the computation across multiple \glspl{gpu} and leveraging mixed-precision arithmetic to optimize accuracy and execution time.
We compare how our eigensolver against state-of-the-art CPU and \gls{fpga} implementations and investigate how mixed-precision enables intermediate operations with higher precision while results are stored with space-efficient representations.

In summary, we present the following contributions: 
\begin{itemize}
    \item A multi-\gls{gpu}, mixed precision, Top-K eigensolver that can process out-of-core sparse matrices with billions of non-zeros. To the best of our knowledge, this is the largest amount reported in the literature (\Cref{sec:implementation}).
    \item A performance evaluation of our \gls{gpu} eigensolver against state-of-the-art Top-K eigensolvers on multiple architectures. We are on average 67$\times$ faster than a multi-core CPU implementation and 1.9$\times$ faster than an \gls{fpga} hardware design, with average reconstruction error below $10^{-5}$ (\Cref{sec:exec_time}).
    \item A characterization of mixed-precision arithmetic in terms of accuracy versus execution time. We prove how mixed-precision is 50\,\% faster than double-precision floating-point arithmetic, and 12$\times$ more accurate than single-precision (\Cref{sec:precision_result}).
\end{itemize}

\begin{figure}[t]
    \centering
    \includegraphics[width=1\columnwidth]{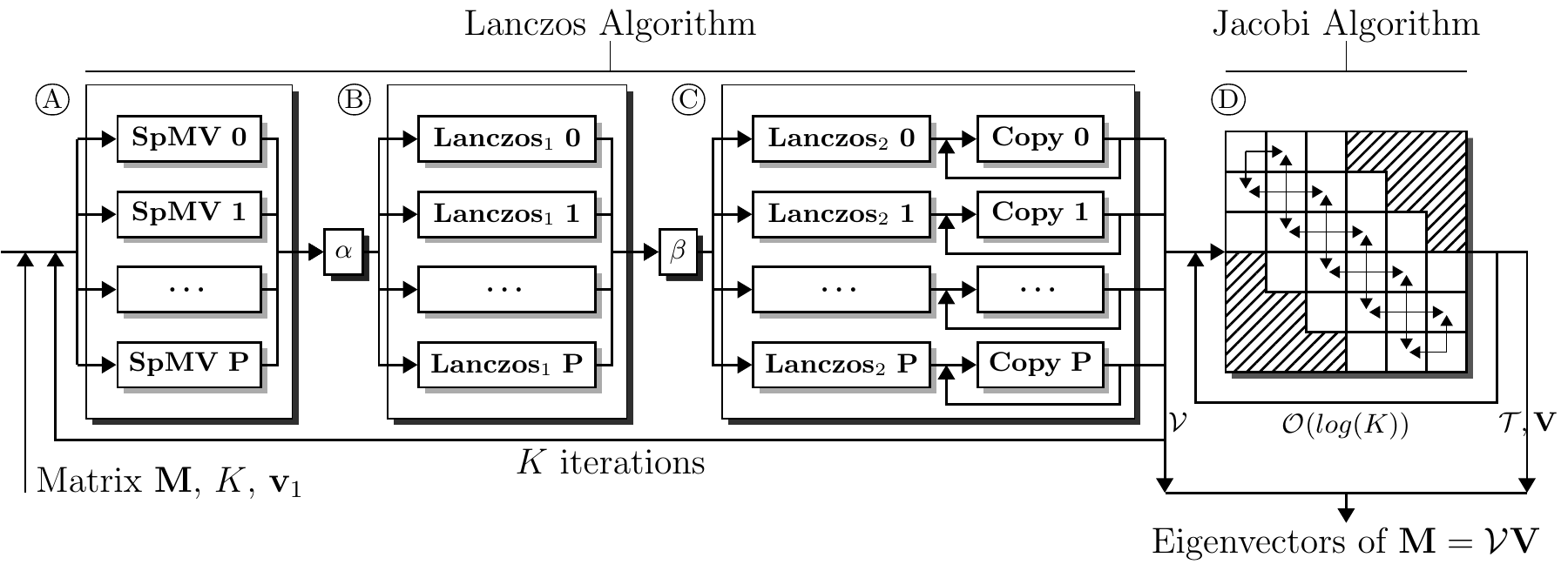}\\
    \caption{High-level design of our Top-K sparse eigensolver, divided into the \lanczos{} and Jacobi algorithms.}
     \label{fig:eigensolver_highlevel}
\end{figure}

\section{Related Work}\label{sec:related}

Although solving Top-K sparse eigenproblems is computationally demanding and has strong practical applications, little prior research optimizes them with hardware accelerators. 
On \glspl{gpu}, most Top-K eigensolvers are domain-specific, do not support large-scale inputs and multiple devices, or are outright not supported on modern \glspl{gpu} architectures ~\cite{aurentz2015gpu,evstigneev2017implementation,dubois2011accelerating}. 
The nvGRAPH library~\cite{nvgraph} by Nvidia uses internally the \lanczos{} algorithm, whose implementation is, however, not user-accessible.
Mixed precision arithmetic on numerical algorithms on \glspl{gpu} has been evaluated on multiple algorithms from the QUDA library. However, the effectiveness of mixed and reduced precision on the numerically unstable \lanczos{} algorithm is still unknown ~\cite{babich2010parallelizing,clark2018pushing}.

Custom hardware designs for Top-K sparse eigensolvers have been recently investigated by Sgherzi et al.~\cite{sgherzi2021solving}, who prototyped their work on high-end \glspl{fpga} equipped with \gls{hbm}.
They investigates the role of mixed-precision arithmetic for Top-K sparse eigensolvers, but the proposed design does not scale to multiple devices or large out-of-core matrices. 
Moreover, limitations of the \gls{fpga}'s \gls{hbm} controller force unnecessary data replication and allow achieving only a fraction of the maximum \gls{hbm} bandwidth. To the best of our knowledge, no other work optimizes large Top-K sparse eigencomputations using \glspl{fpga} or \glspl{dsa}. There are numerous implementations of large-scale Top-K sparse eigenproblem solver for CPUs~\cite{aktulga2012topology,hernandez2009survey,lee2018solution}. However, none is as well-known as ARPACK \cite{lehoucq1998arpack}, a multi-core Fortran library that implements the \gls{iram}, a variation of the \lanczos{} algorithm with support for non-Hermitian matrices.
\section{Implementation}\label{sec:implementation}

Our sparse eigensolver employs a two-phase algorithm, as in \Cref{fig:eigensolver_highlevel}.
The first step is based on the \lanczos{} \cite{lanczos1950iteration} algorithm, which takes as input the original matrix $\mathbf{M}$, the number of desired eigencomponents $K$ and a L2-normalized random vector $v_1 \in \mathbb{R}^{n}$. The algorithm proceeds by incrementally building a Kyrlov subspace (the \lanczos{} vectors $\mathcal{V}$) of $\mathbf{M}$, through vector projections (\Cref{alg:lanczos_low_level} line 9) and orthogonalizations (lines 11--21).
A tridiagonal matrix $\mathcal{T}$ stores the residuals of the previous operations (lines 6, 10) and reduces the problem from size $n \times n$ to one of size $K \times K$ ($K \ll n$).

\algrenewcommand\algorithmicindent{1.2em}%
\begin{algorithm}[t] 
    \caption{Top-K eigenvalues/vectors Lanczos algorithm} \label{alg:lanczos_low_level}
    \begin{algorithmic}[1] 
        
        \Require Input Matrix $\mathbf{M}$, partitioned in $\mathbf{M_1}\ldots\mathbf{M_G}$
        \Require $K$, number of output eigenvectors
        \Require $L2$-normalized input vectors $v_1 := \{v_1^1, \ldots, v_1^G\}$ 
        \Require Temporary vectors $v_{tmp}$ and next vectors $v_{nxt}$
        \Function{Lanczos}{$\mathbf{M},K,v_1,v_{tmp},v_{nxt}$} 
        
        \State $\alpha_1 \gets 0; \quad \beta_1 \gets 0$ \Comment{Initialization}

        \For{$i$ in $1, K$}     
        \LineComment{\textbf{Normalize and compute new Lanczos vector}}
        \If{$i \neq 1$}
            \State $\beta_i \gets \sqrt{\sum_{j = 1}^{G}{(v_{nxt}^j)^2}}$ \Comment{$\beta_i \gets$ $\|v_{nxt}\|_2$}
            \State $v_{i}^{[1\ldots G]} \gets \forall_{j \in [1\ldots G]}(v_{nxt}^j / \beta_i)$ \Comment{$v_i \gets v_{nxt} / \beta_i$}
        \EndIf
        \LineComment{\textbf{Compute the next projection}}
        \State $v_{t}^{[1\ldots G]} \gets SpMV(\mathbf{M_1}\ldots\mathbf{M_G}, v_{i}^{[1\ldots G]})$
        \State $\alpha_i \gets \sum_{j = 1}^{G} v_i^j \cdot  v_t^j$
        \State $v_{nxt}^{[1\ldots G]} \gets v_{tmp}^{[1\ldots G]} - \alpha_i v_{i}^{[1\ldots G]}  - \beta_i v_{i - 1}^{[1\ldots G]}$
        \For{$j \in [1, i]$}
            \Comment{Orthogonalization}
            \If{$j \% 2 \neq 0$}
                \State $o \gets \sum_{k = 1}^{G} v_{j, k} \cdot v_{t, k}$ 
                \State $v_{t}^{[1\ldots G]} \gets v_{t}^{[1\ldots G]} - o \cdot v_{j}^{[1\ldots G]} $
            \Else
                \State $o \gets \sum_{k = 1}^{G} v_{j, k} \cdot v_{n, k}$ 
                \State $v_{n}^{[1\ldots G]} \gets v_{n}^{[1\ldots G]} - o \cdot v_{j}^{[1\ldots G]} $
            \EndIf
            
            \If{$i == j$} \Comment{Copy to temporary vector}
                \State $v_{t}^{[1\ldots G]} \gets v_{n}^{[1\ldots G]}$
            \EndIf
        \EndFor
        \EndFor
        \LineComment{\textbf{Tridiagonal matrix $\mathcal{T}$ and Lanczos vectors $\mathcal{V}$}}
        \State \Return $\{\mathcal{T} = [\alpha_1, \ldots, \alpha_K], [\beta_1, \ldots, \beta_{K - 1}]\}$ 
        \State \Return $\mathcal{V} = [v_1, \ldots, v_K]$
        \EndFunction
    \end{algorithmic}
\end{algorithm}

The second phase employs the Jacobi algorithm \cite{Rutishauser1966jacobi} to solve the eigenproblem on the much smaller matrix $\mathcal{T}$.
The Jacobi algorithm stores the eigenvalues of $\mathbf{M}$ in the main diagonal of $\mathcal{T}$, and the eigenvectors of $\mathcal{T}$ in $\mathbf{V}$.
The eigenvectors of $\mathbf{M}$ are given by $\mathcal{V}\mathbf{V}$.

\subsection{Optimizing sparse eigensolvers for GPUs}\label{sec:optimization}
We desire to render our \gls{gpu} sparse eigensolver scalable to real-world matrices with billions of non-zero values, often encountered in graph analytics.
To do so, we devise a workload partition scheme that distributes the computation across multiple \glspl{gpu} while minimizing unnecessary data movements and synchronization events.

The \lanczos{} algorithm has two synchronization points (\Cref{alg:lanczos_low_level}, lines 6 and 10), corresponding to the computation of $\alpha$ and $\beta$ (\Cref{fig:eigensolver_highlevel} \Circled{A} \Circled{B}). 
Another optional synchronization point occurs if reorthogonalization of the \lanczos{} vectors is needed (lines 15--18, \Cref{fig:eigensolver_highlevel} \Circled{C}).
All other operations operate linearly on the input arrays and can be computed across multiple \glspl{gpu} independently.
The input matrix is partitioned by balancing the number of non-zero elements in each partition.
All vectors, except for $v_i$, are partitioned according to the same partition scheme as the input matrix. 
As the \gls{spmv} performs indirect accesses to the vector $v_i$,
we replicate it to all \gls{gpu} instead of partitioning it.
There is an additional synchronization at each iteration when the previous \lanczos{} vector becomes the input of the \gls{spmv}.
We prevent this synchronization by having each \gls{gpu} copy, in a round-robin fashion, a single partition to a single replica of $v_i$ (\Cref{fig:eigensolver_highlevel} \Circled{C}).
When all \glspl{gpu} have completed a cycle, $v_i$ has been fully copied, and the computation can proceed to a new iteration.

The orthogonality of the \lanczos{} vectors and the quality of the final eigencomponents produced crucially depends on the output of the scalar product ($\alpha$, line 10) and the L2-norm ($\beta$, line 6). 
For this reason, our eigensolver can perform the intermediate operations of each kernel in double-precision floating-point arithmetic to ensure maximum accuracy.
However, vectors can still be stored in single-precision floating-point arithmetic, to consume less device memory and better use the available memory bandwidth.
In our experiments, other data types (half-precision \texttt{FP16}, \texttt{BFLOAT16}) resulted in numerical instability, and have been omitted from \Cref{sec:results}. 


\begin{figure*}[t]
    \centering
    \includegraphics[width=\textwidth]{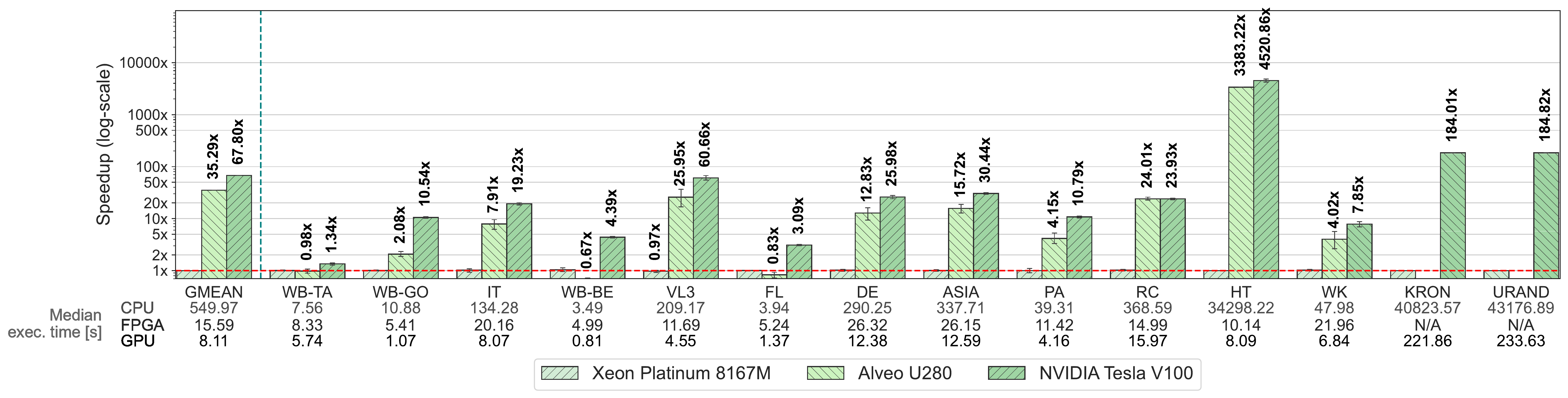}
    \caption{Speedup (log-scale, higher is better) of our GPU Top-K sparse eigensolver versus the ARPACK multi-core CPU library, and the FPGA implementation in Sgherzi et al.~\cite{sgherzi2021solving}. Our GPU implementations runs on a single GPU.
    The two largest matrices (KRON and URAND) are not supported by the FPGA implementation. 
    }
    \label{fig:speedup}
\end{figure*}

\subsection{Implementation details}
We implemented our eigensolver using the GrCUDA API~\cite{parravicini2021dag} and GraalVM~\cite{wurthinger2013one} to support several high-level programming languages automatically, while we wrote the core \gls{gpu} computational kernels in CUDA.
Since GrCUDA internally leverages CUDA unified memory, our eigensolver can scale to out-of-core computations on sparse matrices that would not otherwise fit in the \gls{gpu} memory. 
We also modified the internal GrCUDA runtime to schedule \gls{gpu} kernels across multiple devices, using a round-robin device selection policy for kernels operating on disjoint data.
Through the partition swapping presented in \Cref{sec:optimization}, we minimize unnecessary memory transfers between devices and out-of-core memory pages, guaranteeing scalability in what would be an otherwise memory and transfer-bound computation (\Cref{sec:multigpu_time}). The small tridiagonal matrices that the \lanczos{} algorithm outputs ($\approx 24\times24$) cannot saturate the stream processors of a modern \gls{gpu}~\cite{cosnau2014gpueigen}.
Instead, we achieve better execution time by performing this step on a CPU (\Cref{fig:eigensolver_highlevel} \Circled{D}).

\renewcommand\theadalign{tl}
\renewcommand\theadfont{\bfseries}

\begin{table}
\centering
    \caption{Sparse matrices used in our evaluation, by increasing number of non-zero entries (in millions). We also report the memory footprint in \GB{} of each matrix, stored as COO.}
    \resizebox{1\columnwidth}{!}{
	\begin{tabular}{@{}llllll@{}}
		\toprule
		\thead{ID} & \thead{Name} & \thead{Rows (M)}  & \thead{Non-zeros (M)} & \thead{Sparsity (\%)} & \thead{Size (GB)} \\
		\midrule
        \textbf{WB-TA} & wiki-Talk & 2.39 & 5.02 & $\SI{8.79e-04}{}$ & $\SI{0.06}{\giga\byte}$ \\
        \textbf{WB-GO} & web-Google & 0.91 & 5.11 & $\SI{6.17e-04}{}$ & $\SI{0.07}{\giga\byte}$ \\
        \textbf{WB-BE} & web-Berkstan & 0.69 & 7.60 & $\SI{1.60e-03}{}$ & $\SI{0.10}{\giga\byte}$ \\
		\textbf{FL} & Flickr & 0.82 & 9.84 & $\SI{1.46e-03}{}$ & $\SI{0.13}{\giga\byte}$ \\
		\textbf{IT} & italy\_osm & 6.69 & 14.02 & $\SI{3.13e-05}{}$ & $\SI{0.18}{\giga\byte}$ \\
		\textbf{PA} & patents & 3.77 & 14.97 & $\SI{1.05e-04}{}$ & $\SI{0.19}{\giga\byte}$ \\
		\textbf{VL3} & venturiLevel3 & 4.02 & 16.10 & $\SI{9.96e-05}{}$ & $\SI{0.21}{\giga\byte}$ \\
		\textbf{DE} & germany\_osm & 11.54 & 24.73 & $\SI{1.86e-05}{}$ & $\SI{0.32}{\giga\byte}$ \\
		\textbf{ASIA} & asia\_osm & 11.95 & 25.42 & $\SI{1.78e-05}{}$ & $\SI{0.33}{\giga\byte}$ \\
		\textbf{RC} & road\_central & 14.08 & 33.87 & $\SI{1.71e-05}{}$ & $\SI{0.43}{\giga\byte}$ \\
		\textbf{WK} & Wikipedia & 3.56 & 45.00 & $\SI{3.55e-4}{}$ &  $\SI{0.60}{\giga\byte}$ \\
		\textbf{HT} & hugetrace-00020 & 16.00 & 47.80 & $\SI{1.87e-05}{}$ & $\SI{0.61}{\giga\byte}$ \\
		\textbf{WB} & wb-edu & 9.84 & 57.15 & $\SI{5.90e-05}{}$ & $\SI{0.73}{\giga\byte}$ \\
		\textbf{KRON} & GAP-kron & 134.21 & 4223.26 & $\SI{2.34e-05}{}$ & $\SI{50.67}{\giga\byte}$ \\
		\textbf{URAND} & GAP-urand & 134.21 & 4294.96 & $\SI{2.39e-05}{}$ & $\SI{51.54}{\giga\byte}$ \\
		
		
		


		\bottomrule
	\end{tabular}
    }
    \label{tab:matrices}
\end{table}

\section{Experimental Evaluation}\label{sec:results}

We evaluate the quality of our sparse eigensolver in terms of execution time and results' quality, and provide a performance characterization against state-of-the-art sparse eigensolvers running on different hardware architectures.
We provide an in-depth evaluation over a single \gls{gpu}, and validate the scalability of our algorithm over multiple \glspl{gpu} (up to 8).
Most importantly, we assess the impact of mixed-precision arithmetic and prove how reduced precision results in faster execution time with no meaningful detriment to accuracy.

\subsection{Experimental Setup}\label{sec:setup}
We measure results for our Top-K sparse eigensolver using up to 8 Nvidia Tesla V100s (\GB{16} of HBM2 for each \gls{gpu}).
As baselines, we employ the multi-threaded ARPACK library~\cite{lehoucq1998arpack}, a Top-K sparse eigensolver that uses the \gls{iram} algorithm, running on two Intel Xeon Platinum 8167M (104 threads in total) and \GB{755} of DDR4 memory, with single-precision floating point arithmetic.
We also compare against the recent \gls{fpga} implementation by Sgherzi et al.~\cite{sgherzi2021solving}, running on a Xilinx Alveo U280 accelerator card equipped with \GB{8} of HBM2 memory. 
We repeat measurements 20 times, using random initialization for the \lanczos{} vectors $v_1$.

To provide a fair comparison, we use the same collection of sparse matrices in Sgherzi et al.~\cite{sgherzi2021solving}, enriched with two extremely large matrices (billions of non-zero entries) that do not fit in the \gls{fpga}'s and \gls{gpu}'s device memory, and allows us to test the out-of-core performance of our \gls{gpu} implementation.  
All matrices come from the SuiteSparse collection~\cite{davis2011university} and represent graph topologies, although the eigensolvers in our analysis can be applied to other domains as well~\cite{tung2010enabling}.

\subsection{Execution Time Comparison}\label{sec:exec_time}

We first compare the speed of our \gls{gpu} eigensolver, when running on a single \gls{gpu}, against the CPU and \gls{fpga} baselines, on matrices of increasing size (\Cref{fig:speedup}).
Results have been aggregated over an increasing amount of eigenvectors $K$, from 8 to 24, as the execution time scales linearly with $K$.
For the \gls{fpga} implementation, we use the values reported by the authors. Results of the two largest matrices (KRON and URAND) have been omitted for the \gls{fpga} hardware design as it does not support out-of-core computations.

CPU and \gls{gpu} use single-precision floating point arithmetic, while the \gls{fpga} hardware design uses 32-bit signed fixed point arithmetic with one bit of integer part (\texttt{S1.1.30}) for \lanczos{}, and half-precision floating point arithmetic for Jacobi.

Our \gls{gpu} eigensolver is always faster than both the CPU and \gls{fpga} baselines (on the RC matrix the difference is not statistically significant).
On average, the \gls{gpu} eigensolver is 67$\times$ faster than the CPU implementation and 1.9$\times$ faster than the \gls{fpga} hardware design.
The \gls{fpga} hardware design is still competitive in terms of Performance/Watt, as the \gls{fpga} design consumes 38W~\cite{sgherzi2021solving}, versus the 300W of the \gls{gpu}~\cite{v100}.

As our partitioning minimizes inter-\gls{gpu} data-transfer (\Cref{sec:optimization}), we are $\approx$\,180$\times$ faster than the CPU on very large out-of-core matrices despite storing only a small fraction of the input data on the \gls{gpu} at any given time. 

\begin{figure}%
\centering
\subfigure[]{%
\label{fig:gpu_scaling}%
\includegraphics[width=0.48\columnwidth]{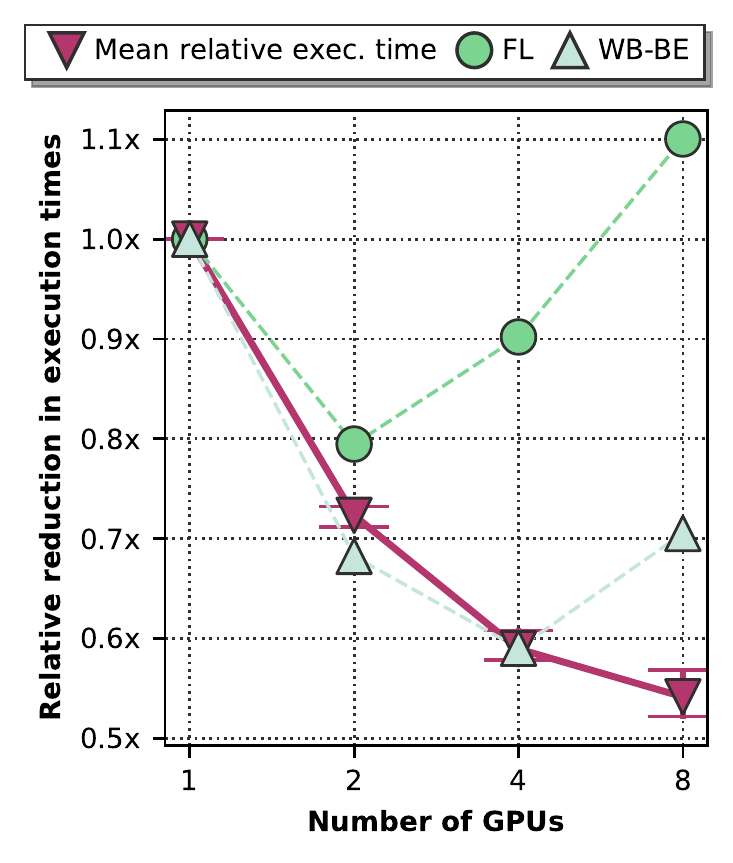}}%
\hfill
\subfigure[]{%
\label{fig:orth_rec}%
\includegraphics[width=0.49\columnwidth]{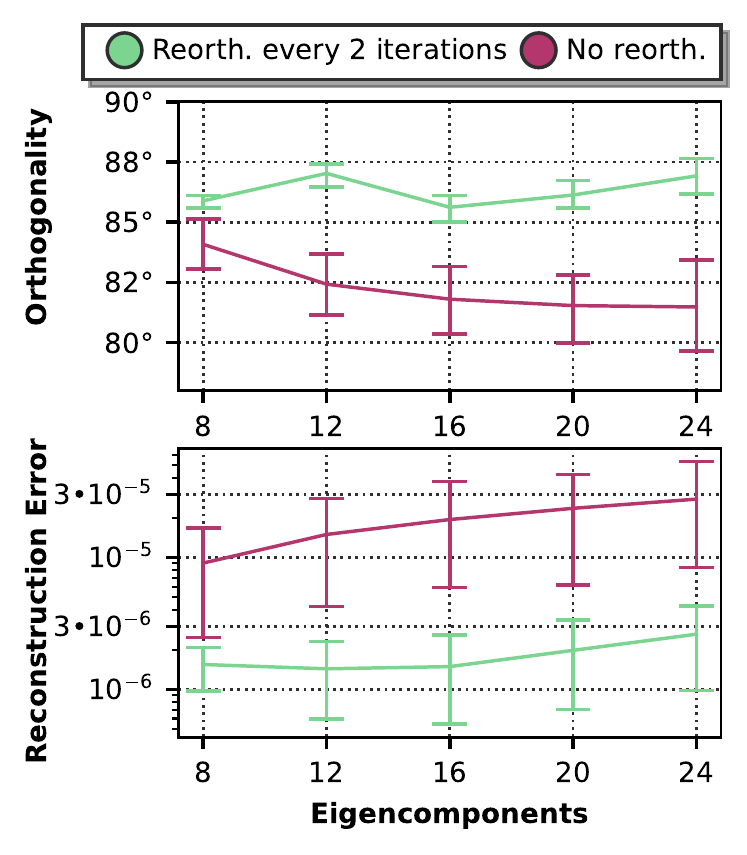}}%
\caption{\Circled{A} Relative execution time for increasing number of \glspl{gpu}. 
    Two outlier matrices (plotted separately) perform worse with more \glspl{gpu} due to the larger inter-\gls{gpu} communication. \Circled{B} Accuracy of our eigensolver, in terms of orthogonality of the eigenvectors and L2 error, for increasing $K$.}
\end{figure}

\subsection{Multi-GPU Performance}\label{sec:multigpu_time}

Scaling the computation of the Top-K eigenvectors on sparse matrices to multiple \glspl{gpu} is far from trivial, as explained in \Cref{sec:optimization}.
From \Cref{fig:gpu_scaling}, we observe how our partitioning scheme improves the execution time when using multiple \glspl{gpu}, with somewhat diminishing returns.
On average, two \glspl{gpu} provide a 50\,\% speedup, while eight \glspl{gpu} are close to a 100\,\% speedup.
Only on two very small matrices we observe a loss of performance on systems with 4 or 8 \glspl{gpu}. 
This phenomenon is explained by the heterogeneous NVLink interconnection found in V100-based systems like ours~\cite{li2019evaluating}. 
Some \gls{gpu} pairs are not directly connected with NVLink, and data transfer has to go through the CPU and PCIe, which has $\approx$\,10$\times$ lower bandwidth than NVLink.

\begin{figure}[t]
    \centering
    \includegraphics[width=0.9\columnwidth]{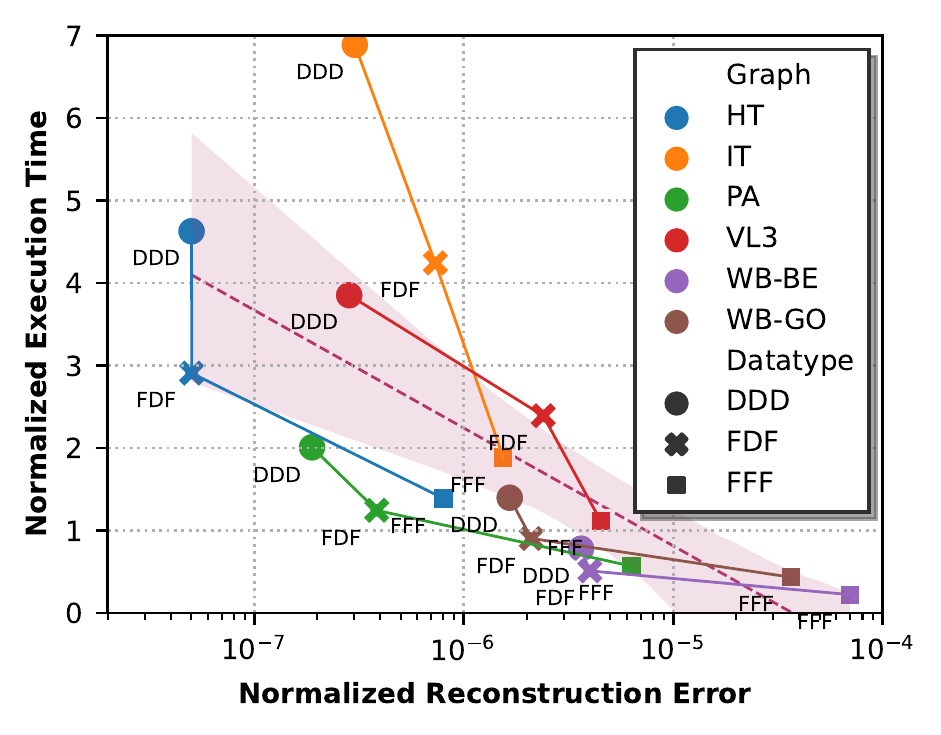}
    \caption{L2 reconstruction error vs. execution time, divided by graph and data-type configuration. The mixed-precision float-double-float configuration is 50\,\% faster than double-precision and has 12$\times$ lower error than pure single-precision. }
    \label{fig:pareto}
\end{figure}

\subsection{Impact of Reorthogonalization and Mixed-precision}\label{sec:precision_result}

To measure the quality of our eigensolver, we measure the average angle that occurs between every pair of eigenvectors.
Eigenvectors are by definition pairwise orthogonal, i.e. their angle is $\pi / 2$, and their dot product must be 0.
\Cref{fig:orth_rec} provides, for increasing $K$, the average orthogonality and the L2 norm of $\mathbf{M}v - \lambda v$, the reconstruction error computed using the definition of eigenvalues.
Both results are aggregated for all matrices due to space limitations. 
We observe how reorthogonalization improves the results' quality, with $\approx$\,2 degrees of difference compared to the implementation without reorthogonalization.
Choosing whether reorthogonalization is suitable or not depends on the application. Spectral methods in machine learning often do not demand the same numerical accuracy as engineering applications, and reorthogonalization increases the algorithmic complexity by an $\mathcal{O}(nK^2/2)$ factor.

Employing mixed-precision arithmetic in numerical algorithms is usually a matter of trade-offs, with better precision translating to higher execution time.
We visualize this behavior in \Cref{fig:pareto}, showing for each matrix the L2 reconstruction error and the relative execution time, and a linear regression to capture the general trend.
In all cases, increasing precision reduces the error and increases the execution time. 
The float-double-float (FDF) configuration (\Cref{sec:optimization}) is 50\,\% faster than a pure double-precision implementation (DDD). Its error is only 40\,\% higher, and 12$\times$ lower than the floating-point implementation (FFF), showing how mixed-precision arithmetic is a great compromise in Top-K sparse eigensolvers.
\section{Conclusion}\label{sec:conclusion}
As graph analytics and spectral methods deal with larger and larger sparse matrices, it is critical to have high-performance Top-K sparse eigensolvers to extract low-dimensional representations of sparse datasets.
We provide a novel \gls{gpu} Top-K sparse eigensolver that can scale to out-of-core matrices with billion of non-zero entries, partition the computation over multiple \glspl{gpu}, and leverage mixed-precision floating-point arithmetic.
We are on average 67$\times$ faster than the multi-core ARPACK CPU library implementation and 1.9$\times$ faster than a state-of-the-art \gls{fpga} hardware design. 
As future work, we will extend our implementation to fixed-point arithmetic and validate if novel interconnection technologies such as NVSwitch can improve even further multi-\gls{gpu} scaling.

\bibliographystyle{IEEEtran}

\IEEEtriggeratref{14}
\bibliography{bibfile}

\begin{thebibliography}{10}
\providecommand{\url}[1]{#1}
\csname url@samestyle\endcsname
\providecommand{\newblock}{\relax}
\providecommand{\bibinfo}[2]{#2}
\providecommand{\BIBentrySTDinterwordspacing}{\spaceskip=0pt\relax}
\providecommand{\BIBentryALTinterwordstretchfactor}{4}
\providecommand{\BIBentryALTinterwordspacing}{\spaceskip=\fontdimen2\font plus
\BIBentryALTinterwordstretchfactor\fontdimen3\font minus
  \fontdimen4\font\relax}
\providecommand{\BIBforeignlanguage}[2]{{%
\expandafter\ifx\csname l@#1\endcsname\relax
\typeout{** WARNING: IEEEtran.bst: No hyphenation pattern has been}%
\typeout{** loaded for the language `#1'. Using the pattern for}%
\typeout{** the default language instead.}%
\else
\language=\csname l@#1\endcsname
\fi
#2}}
\providecommand{\BIBdecl}{\relax}
\BIBdecl

\bibitem{clark2010solving}
M.~A. Clark, R.~Babich, K.~Barros, R.~C. Brower, and C.~Rebbi, ``Solving
  lattice qcd systems of equations using mixed precision solvers on gpus,''
  \emph{Computer Physics Communications}, vol. 181, no.~9, pp. 1517--1528,
  2010.

\bibitem{gupta2015deep}
S.~Gupta, A.~Agrawal, K.~Gopalakrishnan, and P.~Narayanan, ``Deep learning with
  limited numerical precision,'' in \emph{International conference on machine
  learning}.\hskip 1em plus 0.5em minus 0.4em\relax PMLR, 2015, pp. 1737--1746.

\bibitem{clark2018pushing}
M.~A. Clark, A.~Strelchenko, A.~Vaquero, M.~Wagner, and E.~Weinberg, ``Pushing
  memory bandwidth limitations through efficient implementations of
  block-krylov space solvers on gpus,'' \emph{Computer Physics Communications},
  vol. 233, pp. 29--40, 2018.

\bibitem{sze2020efficient}
V.~Sze, Y.-H. Chen, T.-J. Yang, and J.~S. Emer, ``Efficient processing of deep
  neural networks,'' \emph{Synthesis Lectures on Computer Architecture},
  vol.~15, no.~2, pp. 1--341, 2020.

\bibitem{parravicini2020reduced}
A.~Parravicini, F.~Sgherzi, and M.~D. Santambrogio, ``A reduced-precision
  streaming spmv architecture for personalized pagerank on fpga,'' \emph{arXiv
  preprint arXiv:2009.10443}, 2020.

\bibitem{sgherzi2021solving}
F.~Sgherzi, A.~Parravicini, M.~Siracusa, and M.~D. Santambrogio, ``Solving
  large top-k graph eigenproblems with a memory and compute-optimized fpga
  design,'' in \emph{2021 IEEE 29th Annual International Symposium on
  Field-Programmable Custom Computing Machines (FCCM)}.\hskip 1em plus 0.5em
  minus 0.4em\relax IEEE, 2021, pp. 78--87.

\bibitem{ng2002spectral}
A.~Y. Ng, M.~I. Jordan, and Y.~Weiss, ``On spectral clustering: Analysis and an
  algorithm,'' in \emph{Advances in neural information processing systems},
  2002, pp. 849--856.

\bibitem{langville2005survey}
A.~N. Langville and C.~D. Meyer, ``A survey of eigenvector methods for web
  information retrieval,'' \emph{SIAM review}, vol.~47, no.~1, pp. 135--161,
  2005.

\bibitem{dawson2021opinionrank}
G.~Dawson and R.~Polikar, ``Opinionrank: Extracting ground truth labels from
  unreliable expert opinions with graph-based spectral ranking,'' \emph{arXiv
  preprint arXiv:2102.05884}, 2021.

\bibitem{aurentz2015gpu}
J.~L. Aurentz, V.~Kalantzis, and Y.~Saad, ``A gpu implementation of the
  filtered lanczos procedure,'' Tech. Report ys-2015-4, Dept. Computer Science
  and Engineering, University, Tech. Rep., 2015.

\bibitem{evstigneev2017implementation}
N.~M. Evstigneev, ``Implementation of implicitly restarted arnoldi method on
  multigpu architecture with application to fluid dynamics problems,'' in
  \emph{International Conference on Parallel Computational Technologies}.\hskip
  1em plus 0.5em minus 0.4em\relax Springer, 2017, pp. 301--316.

\bibitem{dubois2011accelerating}
J.~Dubois, C.~Calvin, and S.~Petiton, ``Accelerating the explicitly restarted
  arnoldi method with gpus using an autotuned matrix vector product,''
  \emph{SIAM Journal on Scientific Computing}, 2011.

\bibitem{nvgraph}
\BIBentryALTinterwordspacing
Nvidia, ``nvgraph,'' 2019. [Online]. Available:
  \url{docs.nvidia.com/cuda/nvgraph/index.html}
\BIBentrySTDinterwordspacing

\bibitem{babich2010parallelizing}
R.~Babich, M.~A. Clark, and B.~Jo{\'o}, ``Parallelizing the quda library for
  multi-gpu calculations in lattice quantum chromodynamics,'' in \emph{SC'10:
  Proceedings of the 2010 ACM/IEEE International Conference for High
  Performance Computing, Networking, Storage and Analysis}.\hskip 1em plus
  0.5em minus 0.4em\relax IEEE, 2010, pp. 1--11.

\bibitem{aktulga2012topology}
H.~M. Aktulga, C.~Yang, E.~G. Ng, P.~Maris, and J.~P. Vary, ``Topology-aware
  mappings for large-scale eigenvalue problems,'' in \emph{European Conference
  on Parallel Processing}.\hskip 1em plus 0.5em minus 0.4em\relax Springer,
  2012, pp. 830--842.

\bibitem{hernandez2009survey}
V.~Hernandez, J.~Roman, A.~Tomas, and V.~Vidal, ``A survey of software for
  sparse eigenvalue problems,'' \emph{Universitat Politecnica De Valencia,
  SLEPs technical report STR-6}, 2009.

\bibitem{lee2018solution}
D.~Lee, T.~Hoshi, T.~Sogabe, Y.~Miyatake, and S.-L. Zhang, ``Solution of the
  k-th eigenvalue problem in large-scale electronic structure calculations,''
  \emph{Journal of Computational Physics}, vol. 371, 2018.

\bibitem{lehoucq1998arpack}
R.~B. Lehoucq, D.~C. Sorensen, and C.~Yang, \emph{ARPACK users' guide: solution
  of large-scale eigenvalue problems with implicitly restarted Arnoldi
  methods}.\hskip 1em plus 0.5em minus 0.4em\relax SIAM, 1998.

\bibitem{lanczos1950iteration}
C.~Lanczos, \emph{An iteration method for the solution of the eigenvalue
  problem of linear differential and integral operators}, 1950.

\bibitem{Rutishauser1966jacobi}
\BIBentryALTinterwordspacing
H.~Rutishauser, ``The jacobi method for real symmetric matrices,''
  \emph{Numerische Mathematik}, vol.~9, no.~1, pp. 1--10, Nov 1966. [Online].
  Available: \url{https://doi.org/10.1007/BF02165223}
\BIBentrySTDinterwordspacing

\bibitem{parravicini2021dag}
A.~Parravicini, A.~Delamare, M.~Arnaboldi, and M.~D. Santambrogio, ``Dag-based
  scheduling with resource sharing for multi-task applications in a polyglot
  gpu runtime,'' in \emph{2021 IEEE International Parallel and Distributed
  Processing Symposium (IPDPS)}.\hskip 1em plus 0.5em minus 0.4em\relax IEEE,
  2021, pp. 111--120.

\bibitem{wurthinger2013one}
T.~W{\"u}rthinger, C.~Wimmer, A.~W{\"o}{\ss}, L.~Stadler, G.~Duboscq, C.~Humer,
  G.~Richards, D.~Simon, and M.~Wolczko, ``One vm to rule them all,'' in
  \emph{Proceedings of the 2013 ACM international symposium on New ideas, new
  paradigms, and reflections on programming \& software}, 2013, pp. 187--204.

\bibitem{cosnau2014gpueigen}
A.~Cosnuau, ``Computation on gpu of eigenvalues and eigenvectors of a large
  number of small hermitian matrices,'' \emph{Procedia Computer Science},
  vol.~29, pp. 800--810, 12 2014.

\bibitem{davis2011university}
\BIBentryALTinterwordspacing
T.~A. Davis and Y.~Hu, ``The university of florida sparse matrix collection,''
  \emph{ACM Transactions on Mathematical Software (TOMS)}, vol.~38, no.~1, pp.
  1--25, 2011. [Online]. Available: \url{sparse.tamu.edu}
\BIBentrySTDinterwordspacing

\bibitem{tung2010enabling}
F.~Tung, A.~Wong, and D.~A. Clausi, ``Enabling scalable spectral clustering for
  image segmentation,'' \emph{Pattern Recognition}, vol.~43, no.~12, pp.
  4069--4076, 2010.

\bibitem{v100}
Nvidia, ``Nvidia tesla v100 gpu architecture,''
  \url{https://images.nvidia.com/content/volta-architecture/pdf/volta-architecture-whitepaper.pdf},
  2017, retrieved on 2021-10-25.

\bibitem{li2019evaluating}
A.~Li, S.~L. Song, J.~Chen, J.~Li, X.~Liu, N.~R. Tallent, and K.~J. Barker,
  ``Evaluating modern gpu interconnect: Pcie, nvlink, nv-sli, nvswitch and
  gpudirect,'' \emph{IEEE Transactions on Parallel and Distributed Systems},
  vol.~31, no.~1, pp. 94--110, 2019.

\end{thebibliography}
\end{document}